\UseRawInputEncoding
\documentclass[12pt]{iopart}
\expandafter\let\csname equation*\endcsname\relax
\expandafter\let\csname endequation*\endcsname\relax
\usepackage{amssymb, amsmath,color,mciteplus,graphicx,subfigure}
\usepackage{calc,epsfig,epstopdf,color,mciteplus,bm,mathrsfs}
\usepackage{times}
\usepackage{float}
\usepackage{lipsum}
\usepackage{cite}
\usepackage{ulem}
\begin{document}
	
\title{Path-optimized nonadiabatic geometric quantum computation on superconducting  qubits}
	
\author{Cheng-Yun Ding$^1$, Li-Na Ji$^1$, Tao Chen$^{1,*}$ and Zheng-Yuan Xue$^{1,2,3,*}$}

\address{$^1$ Guangdong Provincial Key Laboratory of Quantum Engineering and Quantum Materials, and School of Physics and Telecommunication Engineering, South China Normal University, Guangzhou 510006, China}
\address{$^2$ Guangdong-Hong Kong Joint Laboratory of Quantum Matter, and Frontier Research Institute for Physics, South China Normal University, Guangzhou 510006, China}
\address{$^3$ Guangdong Provincial Key Laboratory of Quantum Science and Engineering, Southern University of Science and Technology, Shenzhen, Guangdong 518055, China}
\address{$^{*}$ Author to whom any correspondence should be addressed.}
\ead{chentamail@163.com and zyxue83@163.com}
	
\date{\today}
	
\begin{abstract}
Quantum computation based on nonadiabatic geometric phases has attracted a broad range of interests, due to its fast manipulation and inherent noise resistance. However,  it is limited to some special evolution paths, and the gate-times are typically longer  than conventional dynamical gates, resulting in weakening of robustness and more infidelities of the implemented geometric gates. Here, we propose a path-optimized scheme for geometric quantum computation on superconducting transmon qubits, where high-fidelity and robust universal nonadiabatic geometric gates can be implemented, based on conventional experimental setups. Specifically, we find that, by selecting appropriate evolution paths, the constructed geometric gates can be superior to their corresponding dynamical ones under different local errors.  Numerical simulations show that   the fidelities for single-qubit geometric Phase, $\pi/8$ and Hadamard gates can be obtained as $99.93\%$, $99.95\%$ and $99.95\%$, respectively. Remarkably, the fidelity for two-qubit control-phase gate can be as high as $99.87\%$. Therefore, our scheme provides a new perspective for geometric quantum computation, making it more promising in the application of large-scale fault-tolerant quantum computation.
\end{abstract}


\maketitle

\section{Introduction}

It is well-known that the physical realization of quantum computers needs a set of universal quantum logic gates, which include a set of arbitrary single-qubit gates and a nontrivial two-qubit gate \cite{Lloyd, Barenco}. However, the coherence of quantum system can be easily ruined by its surrounding environment, invalidating the computation process. Therefore, the fidelity of quantum gates must be high enough so that quantum error correction can be applied to achieve large-scale quantum computation. Moreover, beyond the error-correction threshold, the higher gate-fidelity can result in fewer physical qubit resources encoding an error-free logical qubit. In addition, local random and control errors can also be occurred during the manipulation of the quantum system. Thus, it is also vital to find a robust way of quantum control that is insensitive to these local errors.

One of the promising ways to obtain strong robust quantum gates is suggested to use geometric phase, which was first discovered by Berry \cite{Berry} based on adiabatic process. The geometric phase has global characteristics which only depends on the area of evolution path instead of the evolution details. Thus, quantum gates based on the geometric phase have the intrinsic merit of being robust against local systematic errors, and thereby geometric quantum computation (GQC) \cite{GQC} was received extensively studied with experimentally demonstrations \cite{Jones, berry2, berry3}. However, the required adiabatic evolution condition there \cite{Berry,tong2010} makes operational time of geometric gates to be much longer than that of conventional quantum gates using dynamical evolution, and thus the gate-fidelity is very low in typical quantum systems. In $1987$, to remove the constraint of the adiabatic condition, Aharonov and Anandan \cite{Aharonov} generalize the adiabatic Berry phase to nonadiabatic case. Then, nonadiabatic GQC schemes were proposed \cite{Wang2001,Zhu12002, Zhu2} and verified experimentally \cite{trap, Du, yxu}, which can greatly speed up the geometric gates. Besides, the Berry phase has also been generalized to non-Abelian case \cite{Wilczek}, which also found many applications in quantum computation \cite{Zanardi, Zanardi2, Duan, Tong, Xu}. These advances thus make GQC to be a promising way towards robust quantum manipulation \cite{Chiara,Solinas,Zhu,Filipp,Johansson}.

As we all know, the total phase during quantum evolution consists of both the dynamical and geometric parts. To get a pure geometric phase, the usual method is to eliminate the accompanied dynamical phase, which can be achieved via several special evolution loops \cite{Solinas1,Ota,Thomas}, besides the original time-consuming multi-loop evolution strategies. Therefore, the schemes of GQC based on single-loop evolution have been proposed \cite{Zhang,Xu1,Zhao,Chen,Zhang1,Li2020, Lisai, chentao,cxzhang}, which further decreases the needed time for geometric gates. Notably, the elimination of dynamical phase has also been investigate in the case of quantum computation with non-cyclic geometric phases \cite{Friedenauer2003,Wang12009,Liu,Chen1,duyx, lnji}.

Here, we demonstrate that a given geometric gate can be obtained from various evolution loops that have the different gate robustness. Thus, we take a novel strategy of constructing geometric gates by proposing a path-optimized scheme to realize nonadiabatic GQC and present its implementation on superconducting transmon qubits \cite{Koch,You}. Besides, we take a set of universal single-qubit geometric gates, i.e., the geometric $\pi/2$, $\pi/4$ $Z$-rotation and Hadamard gates, denoted as S, T and H, respectively, as typical examples to test the gate fidelity and robustness. The numerical simulations show that our geometric gates have stronger robustness than the corresponding dynamical gates in certain parametric ranges. Similarly, this path-optimized method can also be readily extended to the case of nontrivial two-qubit geometric gates. Therefore, our scheme provides a new perspective for geometric quantum gates, making them being more promising in future large-scale fault-tolerant quantum computation.

\section{Geometric gates with path optimization}
 \begin{figure}[tbp]\flushright
  \includegraphics[width=0.85\textwidth]{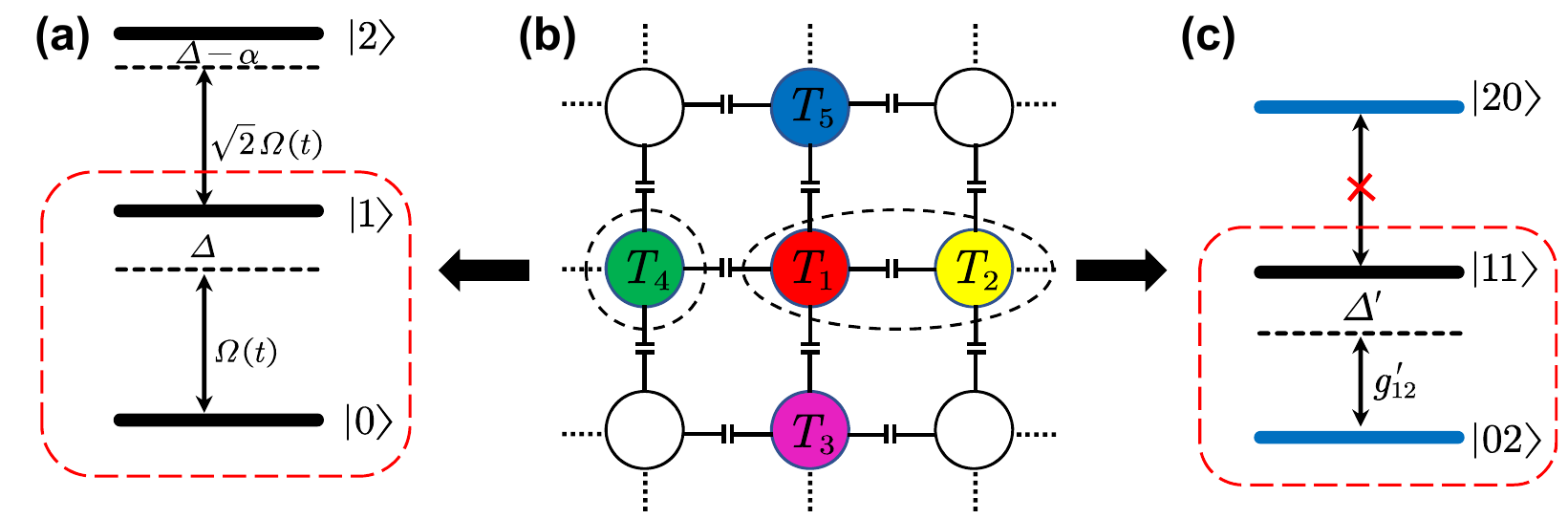}
\caption{Illustration of our implementation. (a) The energy spectrum diagram for a driven transmon qubit $T_4$ with the detuning $\Delta$ and the weak anharmonicity $\alpha$. (b) Schematic diagram of a 2D square transmon lattice linked by coupling capacitively, and the transmon qubits $T_1\sim T_5$ with different colors have different frequencies.  (c) The level structure of second excitation subspace for two parametrically tunable coupled transmons  $T_1$ and $T_2$ with a small detuning $\Delta'$ between energy levels $|11\rangle$ and $|02\rangle$.
  }\label{Figure1}
\end{figure}

In this section, we first present the construction of arbitrary single-qubit nonadiabatic geometric gates on a superconducting transmon qubit with conventional control technology, showing that from our construction, a given geometric gate can be achieved by different evolution paths. Second, we take geometric S, T and H gates as typical examples to test the gate robustness under different evolution paths, which demonstrates that the robustness is different under different paths. Thus, we compare the robustness based on several selected paths with the corresponding dynamical gates, and indicate that our scheme has an advantage range where the robustness of geometric gates is superior to dynamical one. Finally, based on the optimal evolution path, we evaluate these gates performance within actual experimental parameters.

\subsection{The nonadiabatic  geometric phases}
Superconducting quantum circuit system is one of the most promising physical platforms to realize universal quantum computation, due to its flexible controllability and easy scalability. Thanks to the long coherence time, superconducting transmon qubits are usually used to encode quantum information. The energy spectrum for a transmon qubit is shown in figure \ref{Figure1}(a), which has a weak anharmonicity labelled as $\alpha$. For the general requirement of scalable quantum computation, we consider a $2$D square lattice of transmon qubits, where all the adjacent transmons are capacitively coupled, as shown in figure \ref{Figure1}(b). The Hamiltonian of a transmon qubit interacting with an external driving microwave field can be described as
 \begin{eqnarray}\label{Htotal}
\mathcal{H}(t)=\omega_{0}|1\rangle\langle1|+(2\omega_0-\alpha)|2\rangle\langle2|  \nonumber \\
\qquad\quad\;\!+\frac{1}{2}\left\{\Omega(t)e^{i\left[\int_0^t\omega(t')dt'-\phi(t)\right]}
(|0\rangle\langle1|+\sqrt{2}|1\rangle\langle2|)+\textup{H.c}\right\},
\end{eqnarray}
where $\omega_0$ is transition frequency of the transmon qubit with the energy of ground state setting to be zero, $\Omega(t)$, $\omega(t)$ and $\phi(t)$ are driving strength, frequency and phase of microwave field, respectively. Note that we here use $\{|0\rangle,|1\rangle\}$ as our computational subspace, and consider the higher energy level $|2\rangle$ as the main leakage source, due to the weak anharmonicity of transmon qubit. Applying a representation transformation of $U\!=\!U_2\times U_1$ to the Hamiltonian in equation (\ref{Htotal}) with
$U_1\!=\!\exp\left\{-i\omega_0(|1\rangle\langle1|+2|2\rangle\langle2|)t\right\}$, $U_2\!=\!\exp\{i\int_0^t\Delta(t')dt'(|1\rangle\langle1|\!-\!|0\rangle\langle0|+3|2\rangle\langle2|)/2\}$, the transformed Hamiltonian $\mathcal{H}'(t)$ will be
\begin{eqnarray}
\mathcal{H}'(t)=\frac{1}{2}\{\Delta(t)(|1\rangle\langle1|-|0\rangle\langle0|+3|2\rangle\langle2|) -2\alpha|2\rangle\langle2|\} \nonumber\\
\qquad\quad\;+\frac{1}{2}\left\{\Omega(t)e^{-i\phi(t)}(|0\rangle\langle1|+\sqrt{2}|1\rangle\langle2|) +\textup{H.c}\right\},
\end{eqnarray}
where the detuning is defined as $\Delta(t)=\omega_0-\omega(t)$.

Next, we consider the ideal case, i.e., projecting the Hamiltonian $\mathcal{H}'(t)$ within the qubit subspace, to show the construction of arbitrary single-qubit nonadiabatic geometric gates with path optimization. Then, the Hamiltonian $\mathcal{H}'(t)$ can be reduce to
\begin{eqnarray}
\mathcal{H}_1(t)=-\frac{1}{2}\left\{\Delta(t)\sigma_z-\Omega(t) \left[\cos{\phi(t)}\sigma_x+\sin{\phi(t)}\sigma_y\right]\right\},
\end{eqnarray}
with $\{\sigma_x,\sigma_y,\sigma_z\}$ being well-known Pauli operators. We further choose a set of dressed-state bases $\{|\varphi_\pm(t)\rangle\}$ as the evolution states which are, respectively,
\begin{eqnarray}
\begin{split}
|\varphi_{+}(t)\rangle&=e^{i k_{+}(t)}\left[\cos(\chi(t)/2)|0\rangle+\sin(\chi(t)/2)e^{i\xi(t)}|1\rangle\right],  \\
|\varphi_{-}(t)\rangle&=e^{ik_{-}(t)}\left[\sin(\chi(t)/2)e^{-i\xi(t)}|0\rangle -\cos{(\chi(t)/2)}|1\rangle\right],
\end{split}
\end{eqnarray}
where $k_{\pm}(t)$ satisfy $k_{\pm}(0)\!\!=\!\!0$, and their evolution trajectories on the Bloch sphere can be determined by the time-dependent polar angle $\chi(t)$ and azimuth angle $\xi(t)$. By solving the Schr\"{o}dinger equation $i |\dot{\varphi}_\pm(t)\rangle=\mathcal{H}_1(t)|\varphi_\pm(t)\rangle$, we can get the parameters that determine the evolution trajectories as
\begin{eqnarray}\label{pathparameter}
\begin{split}
\dot{\chi}(t)&=\Omega(t)\sin[\phi(t)-\xi(t)],  \\
\dot{\xi}(t)&=-\Delta(t)-\Omega(t)\cot\chi(t)\cos[\phi(t)-\xi(t)],
\end{split}
\end{eqnarray}
as well as
\begin{eqnarray}
k_{+}(t)=-k_{-}(t)=-\frac{1}{2}\int_0^{t}\left\{\dot{\xi}(t') [\cos{\chi(t')}-1]-\Delta(t')\right\}/\cos{\chi(t')}dt'.
\end{eqnarray}
Assuming that the evolution trajectory is determined, which corresponds to the known polar and azimuth angles, then the relevant parameters $\{\Omega(t),\phi(t),\Delta(t)\}$ of Hamiltonian $\mathcal{H}_1(t)$ can be obtained reversely by solving the above restriction relations. Therefore, after a period of cyclic evolution $\tau$, the evolution states will be transformed into $U(\tau)|\varphi_\pm(0)\rangle=e^{i k_{\pm}(\tau)}|\varphi_\pm(0)\rangle$, whose corresponding evolution operator is
\begin{eqnarray}\label{EqU}
\begin{split}
U(\tau)&=e^{i k_{+}(\tau)}|\varphi_+(0)\rangle\langle \varphi_+(0)|+e^{i k_{-}(\tau)}|\varphi_-(0)\rangle\langle \varphi_-(0)|   \\
&=\cos{\gamma}+i\sin{\gamma}\left[\sigma_z\cos{\chi_0}+\sin{\chi_0} (\sigma_x\cos{\xi_0}+\sigma_y\sin{\xi_0})\right] \\
&=e^{i\gamma\:\vec{n}\cdot\vec{\sigma}},
\end{split}
\end{eqnarray}
where $\gamma\!=\!k_{+}(\tau)\!=\!-k_{-}(\tau)\!=\!\gamma_d+\gamma_g$ is the accumulated total phase during the evolution process, $\chi_0\!=\!\chi(0)$ and $\xi_0\!=\!\xi(0)$ are the initial polar and azimuth angles, respectively. Since $\vec{n}=(\sin\chi_0\cos\xi_0, \sin\chi_0\sin\xi_0, \cos\chi_0)$ is an arbitrary unit direction vector and $\vec{\sigma}=(\sigma_x, \sigma_y, \sigma_z)$ is a vector for Pauli operators, $U(\tau)$ represents a rotation along the $\vec{n}$  axis with an angle of $-2\gamma$. In addition, as the dynamical phase $\gamma_d$ can be calculated as
\begin{eqnarray}
\label{dynamical}
\gamma_d&=&-\int^{\tau}_0\langle\varphi_{+}(t)|\mathcal{H}_1(t)|\varphi_{+}(t)\rangle dt  \nonumber \\
&=&\frac{1}{2}\int^{\tau}_0 [\dot{\xi}(t)\sin^2\chi(t)+\Delta(t)]/\cos\chi(t) dt ,
\end{eqnarray}
 the remaining geometric phase $\gamma_g$ will be
\begin{eqnarray}
\label{geometric}
\gamma_g=\gamma-\gamma_d=-\frac {1} {2}\int^{\tau}_0 \dot{\xi}(t)\left[1-\cos\chi(t)\right]dt.
\end{eqnarray}
Quantum logic gates constructed by geometric phases have intrinsic fault-tolerance feature, as the geometric phase  only depend on the global geometric properties of the path, instead of the specific evolution details. This can be explained by the fact that, in equation (\ref{geometric}), the geometric phase is equal to half of the solid angle enclosed by the evolution path. However, the existence of dynamical phase, as shown in equation (\ref{dynamical}), leads to the appearance of a non-geometric component in the total phase, which will destroy the global noise-resistance feature of the constructed geometric gate, and thus need be eliminated to get a pure geometric phase.

\subsection{Construction of path-optimized geometric gates}
In this subsection, we consider the construction of path-optimized geometric gates, with the above proposed method of inducing geometric phases. In the above construction, the detuning is set to be in a  generally time-dependent form, but we now set it as a constant,  following the experimentally preference. That is, we set $\Delta(t)=\Delta$, and consider the case that the closed evolution paths are formed by the longitude and latitude lines, for simplicity.  For the longitude line, $\dot{\xi}(t)=0$ will be always hold, and thus setting $\Delta=0$ will lead to $\gamma_d=0$ in equation (\ref{dynamical}), which also coincidence with the simplest experimental control. In this case, by solving the equation (\ref{pathparameter}), the shapes of parameters $\{\Omega(t),\phi(t)\}$ can be reduced  to
\begin{subequations}\label{para}
\begin{eqnarray}
\Omega(t)=\left\{
            \begin{array}{l}
\dot{\chi}(t),   \quad\phi(t)=\xi(t)+\pi/2;\\
-\dot{\chi}(t),   \quad\phi(t)=\xi(t)-\pi/2.
            \end{array}
          \right.
\end{eqnarray}
And, for the latitude line,  by setting $\gamma_d=0$ in equation (\ref{dynamical}) and letting the detuning to be time-independent for simple experimental control, i.e., $\Delta(t)\equiv \Delta$, we obtain that
\begin{eqnarray}
\Delta=-\frac{1}{\tau}\int^{\tau}_0\dot{\xi}(t)\sin^{2}\chi(t)dt.
\end{eqnarray}
Then, other parameters can be solved from equation (\ref{pathparameter})  as
\begin{eqnarray}
\Omega(t)=\left\{
            \begin{array}{l}
            [\dot{\xi}(t)+\Delta]\tan{\chi(t)}, \quad  \phi(t)=\xi(t)+\pi;  \\
            -[\dot{\xi}(t)+\Delta]\tan{\chi(t)}, \quad \phi(t)=\xi(t).
            \end{array}
          \right.
\end{eqnarray}
\end{subequations}
Note that different selection of the $\pm$ sign and the corresponding expression of $\phi(t)$ is to ensure that the pulse always is positive, which is experimental friendly.
In addition, according to equation (\ref{geometric}), the geometric phase $\gamma_g$ is only related to the polar angle $\chi(t)$ and the azimuth angle $\xi(t)$ of the evolution path, and thus different choices of parameters $\{\chi(t), \xi(t)\}$ may induce the same $\gamma_g$. Specifically, if the rotation axis $\vec{n}$ is determined, i.e., the initial value $(\chi_0, \xi_0)$ of the evolution state is determined, different evolution paths can be chosen for each specific single-qubit geometric gate.

\begin{figure*}[!t]\flushright
\subfigure {\includegraphics[width=0.85\textwidth]{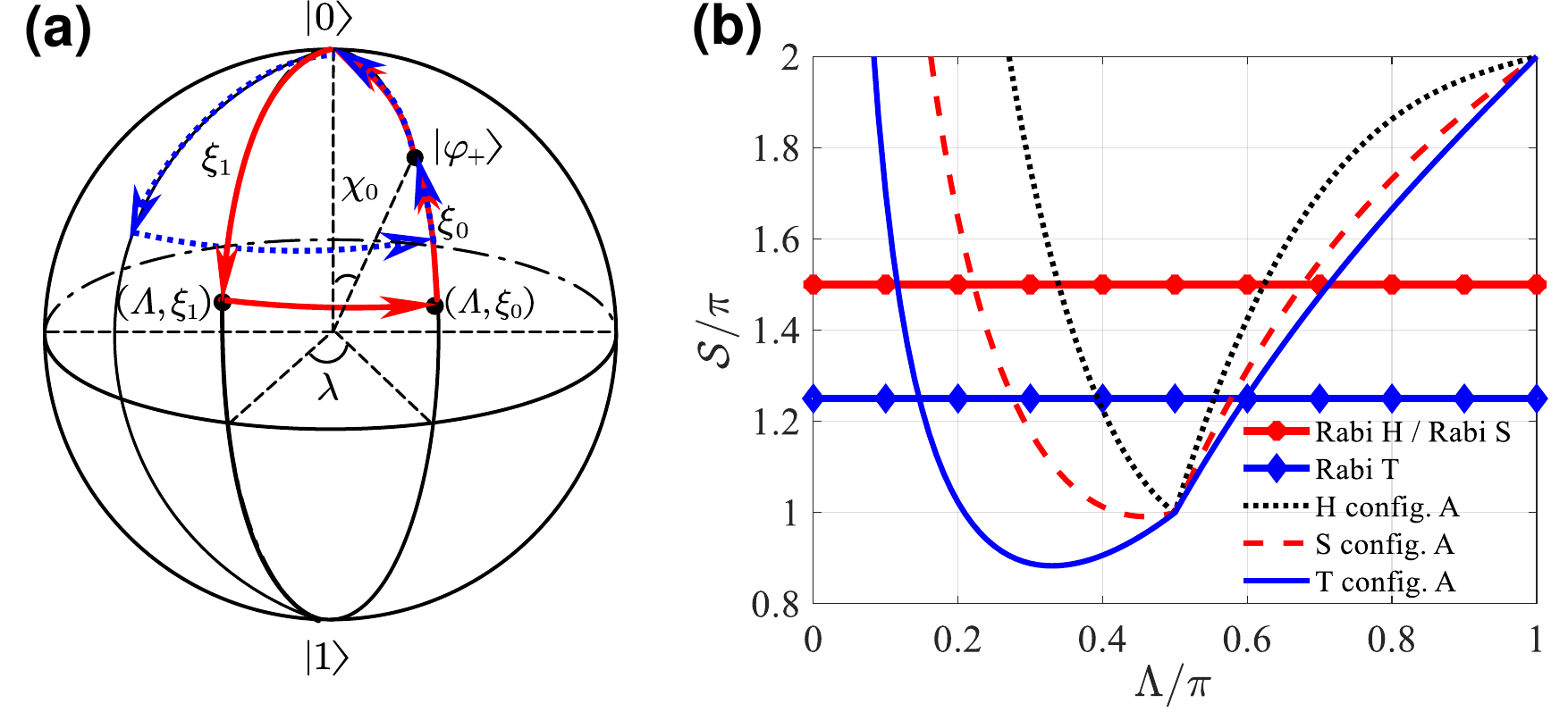}}
\caption{Illustration of our path-optimized single-qubit geometric gates construction (the configuration A) and the needed pulse areas. (a) The closed evolution trajectories  for a gate  in the Bloch sphere, where different colours  denote  trajectories with different path parameters $\Lambda$. (b) The comparison of pulse areas $\mathcal{S}$ between geometric S, T and H gates with different path parameters and their corresponding dynamical gates. }\label{Figure2}
\end{figure*}

Thus, as shown in figure \ref{Figure2}(a), to realize universal single-qubit nonadiabatic geometric gates, the evolution path can be divided into four parts in the Bloch sphere.  To meet the parameters' constraint as described in equation (\ref{para}), the parameters $\{\Omega(t),\phi(t),\Delta(t)\}$ of microwave fields in each part should meet the following conditions,
\begin{subequations}\label{eq_evolution}
\begin{eqnarray}
\mathcal{S}_1=\int^{\tau_1}_0\Omega(t)dt=\chi_0, \quad \Delta=0, \quad  \phi=\xi_0-\frac{\pi}{2};
\end{eqnarray}
\begin{eqnarray}
\mathcal{S}_2=\int^{\tau_2}_{\tau_1}\Omega(t)dt=\Lambda, \quad \Delta=0, \quad \phi=\xi_1+\frac{\pi}{2};
\end{eqnarray}
\begin{eqnarray}
\mathcal{S}_3=\int^{\tau_3}_{\tau_2}\Omega(t)dt=|\lambda\tan\Lambda\cos^2\Lambda|, \quad \Delta=\frac{\lambda\sin^2\Lambda}{\tau_3-\tau_2}, \notag \\
\quad \phi(t)= \left\{
            \begin{array}{l}
\xi(t)+\pi, \quad   0<\Lambda<\pi/2, \\
\xi(t),  \quad  \pi/2<\Lambda\leq\pi,
            \end{array}
          \right.  \quad \int^{\tau_3}_{\tau_2}\dot{\xi}(t)=-\lambda;
\end{eqnarray}
\begin{eqnarray}
\mathcal{S}_4=\int^{\tau}_{\tau_3}\Omega(t)dt=|\Lambda-\chi_0|, \quad \Delta=0, \notag \\
\quad  \phi= \left\{
            \begin{array}{l}
\xi_0-\frac{\pi}{2}, \quad \Lambda>\chi_0, \\
\xi_0+\frac{\pi}{2}, \quad\Lambda<\chi_0.
            \end{array}
          \right.
\end{eqnarray}
\end{subequations}
where $\mathcal{S}_i$ denotes the pulse area of the $i$th segment, $\lambda=\xi_1-\xi_0$, $\Lambda$ equals to value of the polar angle $\chi(t)$ on the third segment ($t\in[\tau_2,\tau_3]$), which is a constant and  $\Lambda\in(0,\pi/2)\cup(\pi/2,\pi]$.  Note that, the  shapes  of $\Omega(t)$ and $\phi(t)$, determined by $\chi(t)$ and $\xi(t$),  prescribed in equation (\ref{eq_evolution}) is still not limited to special cases, the only requirement is that the integrals  must be certain values, and thus various pulse-shaping techniques can be further incorporated into the construction, which is not the topic here and not included.
This choice of $\xi_1$ leads to a anticlockwise trajectory as shown in figure \ref{Figure2}(a), and we term it configuration A.   By this setting, at the final time $\tau$, the evolution operator will be
\begin{eqnarray}\label{eq_universal}
U_1(\tau)=U_1(\tau,\tau_3)U_1(\tau_3,\tau_2)U_1(\tau_2,\tau_1)U_1(\tau_1,0)
=e^{i\gamma_g\vec{n}\cdot\vec{\sigma}},
\end{eqnarray}
which is an arbitrary single-qubit geometric gate and $\gamma_g\!=\!\lambda(1-\cos\Lambda)/2$ (assuming $\gamma_g<0$). Since $\gamma_g$ is only related to $\Lambda$ and $\lambda$, we can choose different evolutionary trajectories to realize the same specific gate under the condition of fixing the evolutionary starting point. For example, shown in figure \ref{Figure2}(a), the evolution state $|\varphi_{+}(t)\rangle$ can return to the starting point along either path $1$  (red solid line)  or path $2$  (blue dotted line)  to get the same geometric gate. Note that paths $1$ and $2$ have different $\Lambda$, so we term it as the ``path parameter''.

Specifically, in order to implement geometric S, T and H gates, we can set
\begin{eqnarray}
\begin{split}
\left(\chi_0,\xi_0\right)&=\left(0,0\right),   \quad\quad\quad\! \gamma_g=-\pi/4,  \\
\left(\chi_0,\xi_0\right)&=\left(0,0\right),   \quad\quad\quad\! \gamma_g=-\pi/8, \\
\left(\chi_0,\xi_0\right)&=\left(\pi/4,0\right),   \quad\;\;\; \gamma_g=-\pi/2,
\end{split}
\end{eqnarray}
respectively. As it is well-known that the gate sequence of them can construct arbitrary single-qubit geometric gates. Moreover, for a certain $\gamma_g$,  different choice of parameters $\{\Lambda, \lambda\}$ corresponds to different evolution paths, with which we can realize the same geometric gate but with different gate-robustness.

In addition, as in figure \ref{Figure2}(b), we also consider the needed gate-time  of these geometric gates with  different  path parameters $\Lambda$. Here, the  gate-time is defined  by their corresponding pulse areas $\mathcal{S}=\sum_{i=1}^4 \mathcal{S}_i=\int_0^\tau\Omega(t)d t$.  Besides, we compare the gate-time in our scheme with the corresponding dynamical one. It can be concluded that geometric S, T and H gates have shorter gate-time than their corresponding dynamical gates within the  path parameter ranges of  $\Lambda/\pi\in[0.23,0.5)\cup(0.5,0.67]$, $[0.15,0.5)\cup(0.5,0.59]$ and $[0.34,0.5)\cup(0.5,0.62]$, respectively.  Significantly, within a fairly wide range of paths, the pulse areas of these gates are less than $2\pi$ that corresponds to those of the conventional nonadiabatic GQC schemes \cite{Zhao,Chen,Zhang1}, which is also one of the merits of our scheme.

\begin{figure}[tbp]
\flushright
  \includegraphics[width=0.85\textwidth]{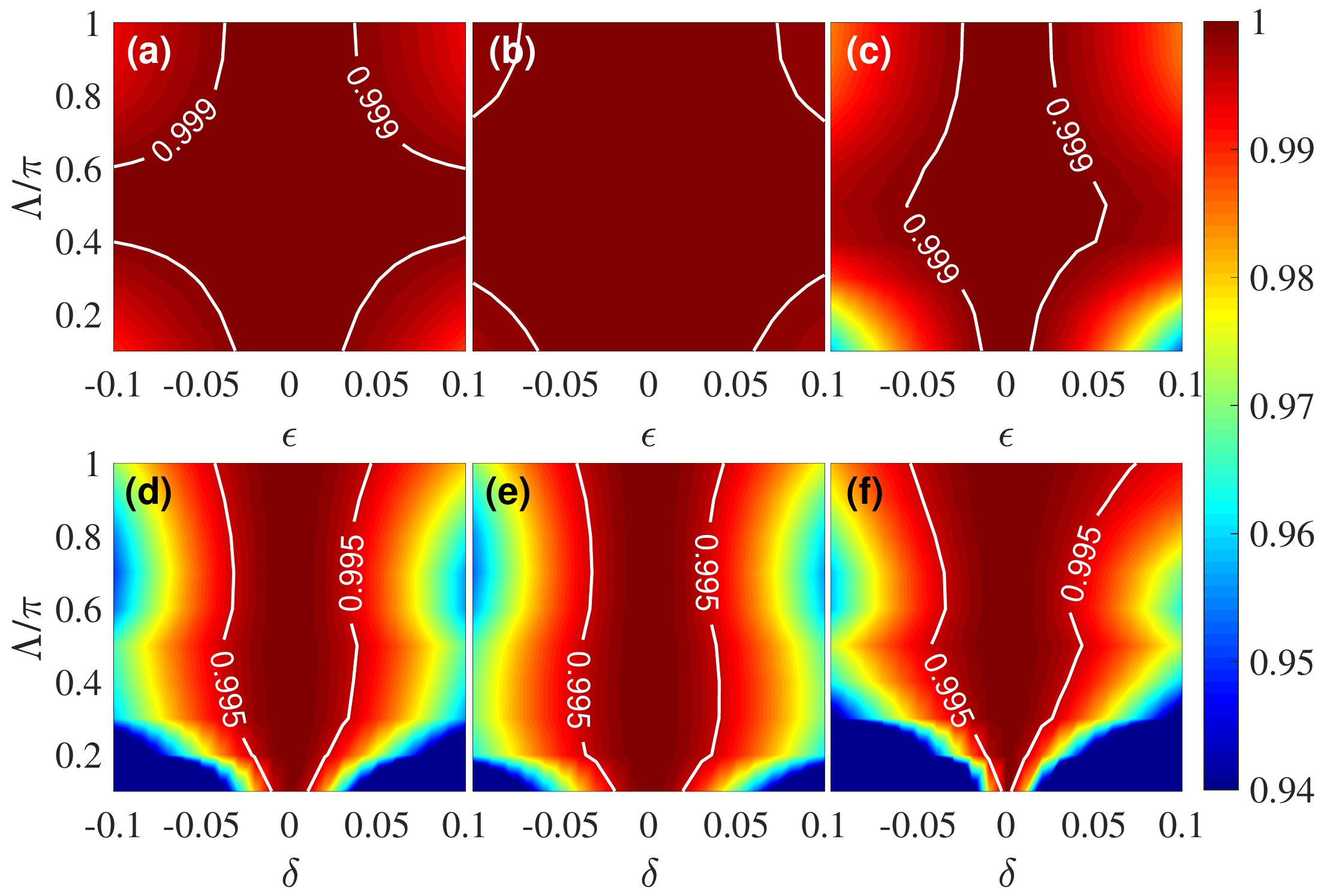}
  \caption{Gate fidelities for geometric S, T and H gates as function of path parameters $\Lambda$, (a-c) $\sigma_x$  and (d-f) $\sigma_z$ errors with error fractions $\epsilon, \delta\in[-0.1,0.1]$,  without decoherence.}\label{Figure3}
\end{figure}

\subsection{Gate robustness}

In the following, we  evaluate the gate robustness of the implemented geometric S, T and H gates with path selection. Under the conventional errors against the amplitude deviation of the driving fields and the qubit-frequency drift, i.e., $\sigma_x$ and $\sigma_z$ errors, the Hamiltonian $\mathcal{H}_1(t)$ will be changed into
\begin{eqnarray}
\mathcal{H}'_1(t)=-\frac{1}{2}\left\{\left[\Delta+\delta\Omega_0\right]\sigma_z-(1+\epsilon)\Omega(t) \left[\cos{\phi(t)}\sigma_x+\sin{\phi(t)}\sigma_y\right]\right\},
\end{eqnarray}
where $\epsilon$ and $\delta$ represent the error fractions for two errors in unit of the time-dependent amplitude $\Omega(t)$ and its maximum amplitude $\Omega_0$, respectively. Besides, we use the definition of gate fidelity, under the two errors \cite{Wang1},
\begin{eqnarray}
F_{G}^{\epsilon,\delta}&=\textrm{Tr}(U_1^{\dagger}U_1^{\epsilon,\delta})/\textrm{Tr}(U_1^{\dagger}U_1),
\end{eqnarray}
to evaluate the gate robustness, where $U_{1}^{\epsilon,\delta}$ and $U_1$ are the evolution  operators with and without errors, respectively. In figure \ref{Figure3}, we show the gate fidelities of the geometric S, T and H gates as function of path parameter $\Lambda/\pi\in[0.1,0.5)\cup(0.5,1]$ and the two kinds of local errors $\epsilon,\delta\in[-0.1,0.1]$. We can clearly see that the gate robustness is different for different  $\Lambda$, and we target to find  ranges of the path  parameter $\Lambda$, where the geometric gate-robustness is better than that of the corresponding dynamical one for both errors.

\begin{figure}[tbp]
\flushright
\includegraphics[width=0.85\linewidth]{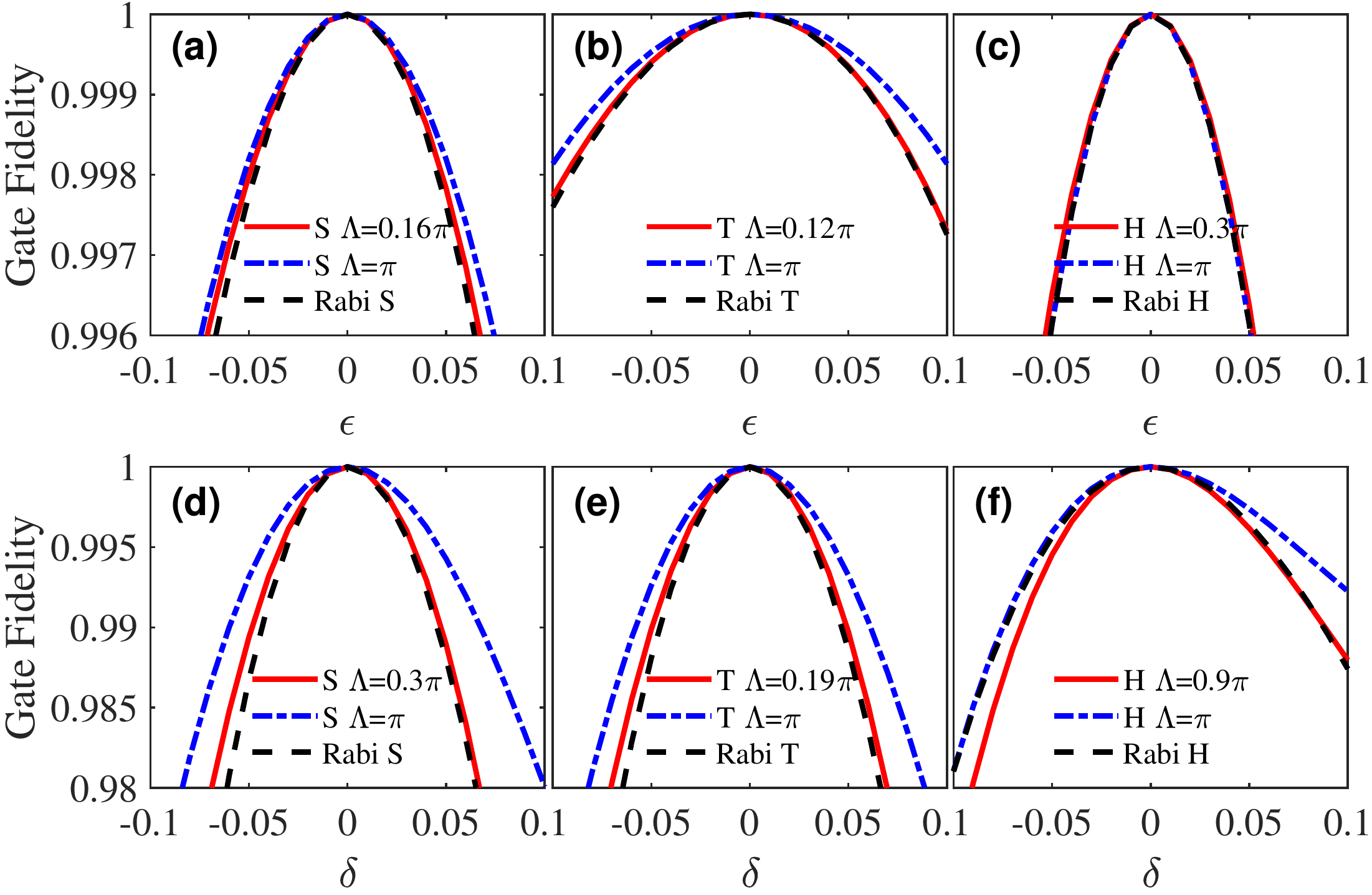}
\caption{The comparison of robustness between geometric S, T and H gates and their corresponding dynamical gates (denoted as ``Rabi gates'') for two local errors, i.e., (a-c) $\sigma_x$ and (d-f) $\sigma_z$ errors, for their boundary paths, respectively.
}\label{Figure4}
\end{figure}

We now compare the robustness between geometric and corresponding dynamical gates (denoted as ``Rabi gates'') against two errors, and the construction of the dynamical gates is detailed in appendix A. In figure \ref{Figure4}, we plot the boundary paths, i.e., the robustness of both strategies is approximately the same. And, within the boundaries, the robustness of the geometric gates is better than the corresponding dynamical ones. Therefore, we can obtain that the robustness of geometric S and T gates is better than that of the corresponding dynamical gates against both errors within ranges of $\Lambda/\pi\in[0.3,0.5)\cup(0.5,1]$ and $[0.19,0.5)\cup(0.5,1]$, respectively. For the geometric H gate, its robustness is better than that of the corresponding dynamical gate against $\sigma_x$ error in the range $\Lambda/\pi\in[0.3,0.5)\cup(0.5,1)$ and that against $\sigma_z$ error only for the path $\Lambda/\pi=1$, which reduces to the orange-slice-shaped-loop case \cite{Zhao, Chen, Zhang1}.  Remarkably, there is another kind of path configuration, we term it as the ``configuration B'', in contrast to the ``configuration A'' of described above. We find that, for the case of H gate against $\sigma_z$ error, there is still a parameter range of $\Lambda/\pi\in[0.65,0.75]$,  within which geometric H gate has better robustness than the corresponding dynamical gate, and the details are presented in appendix B. So, for quantum systems with  different dominant errors, we can utilize the different  configuration strategies to construct geometric H gate with strong robustness, extending the robust characteristic  for  the orange-slice-shape scheme \cite{yxu} to our shorter path scheme.

\subsection{Gate performance on a transmon under decoherence}
In practical superconducting system, due to the weak anharmonicity of a  transmon qubit, the state population in computational subspace $\{|0\rangle,|1\rangle\}$ will leak to higher levels, e.g. $|2\rangle$, which leads to the decrease of fidelity. Thus, it is necessary to use ``derivative removal via adiabatic gate''  technique \cite{Motzoi,Gambetta} to correct this leakage. Here, we can simulate numerically the gate fidelity by setting  appropriate  parameters, using the Lindblad master equation \cite{Lindblad} of
\begin{eqnarray}\label{master}
\dot{\rho}_1=-i[\mathcal{H}(t),\rho_1]+\frac{1}{2}\left\{\kappa_z\mathcal{L}(\mathcal{A}_z) +\kappa_{-}\mathcal{L}(\mathcal{A}_{-})\right\},
\end{eqnarray}
where $\rho_1$ is the density matrix  of the quantum system and its quantum dynamics is governed  by the Hamiltonian $\mathcal{H}(t)$, $\mathcal{L}(\mathcal{A})=2\mathcal{A}\rho_1\mathcal{A}^{\dagger} -\mathcal{A}^{\dagger}\mathcal{A}\rho_1-\rho_1\mathcal{A}^{\dagger}\mathcal{A}$ is the Lindblad operator for decay operator $\mathcal{A}_{-}=|0\rangle\langle1|+\sqrt{2}|1\rangle\langle2|$ and dephasing operator $\mathcal{A}_z=|1\rangle\langle1|+2|2\rangle\langle2|$, and $\kappa_{-}$, $\kappa_z$ are the relaxation and dephasing rates, respectively.

\begin{figure}[tbp]
\flushright
\includegraphics[width=0.7\textwidth]{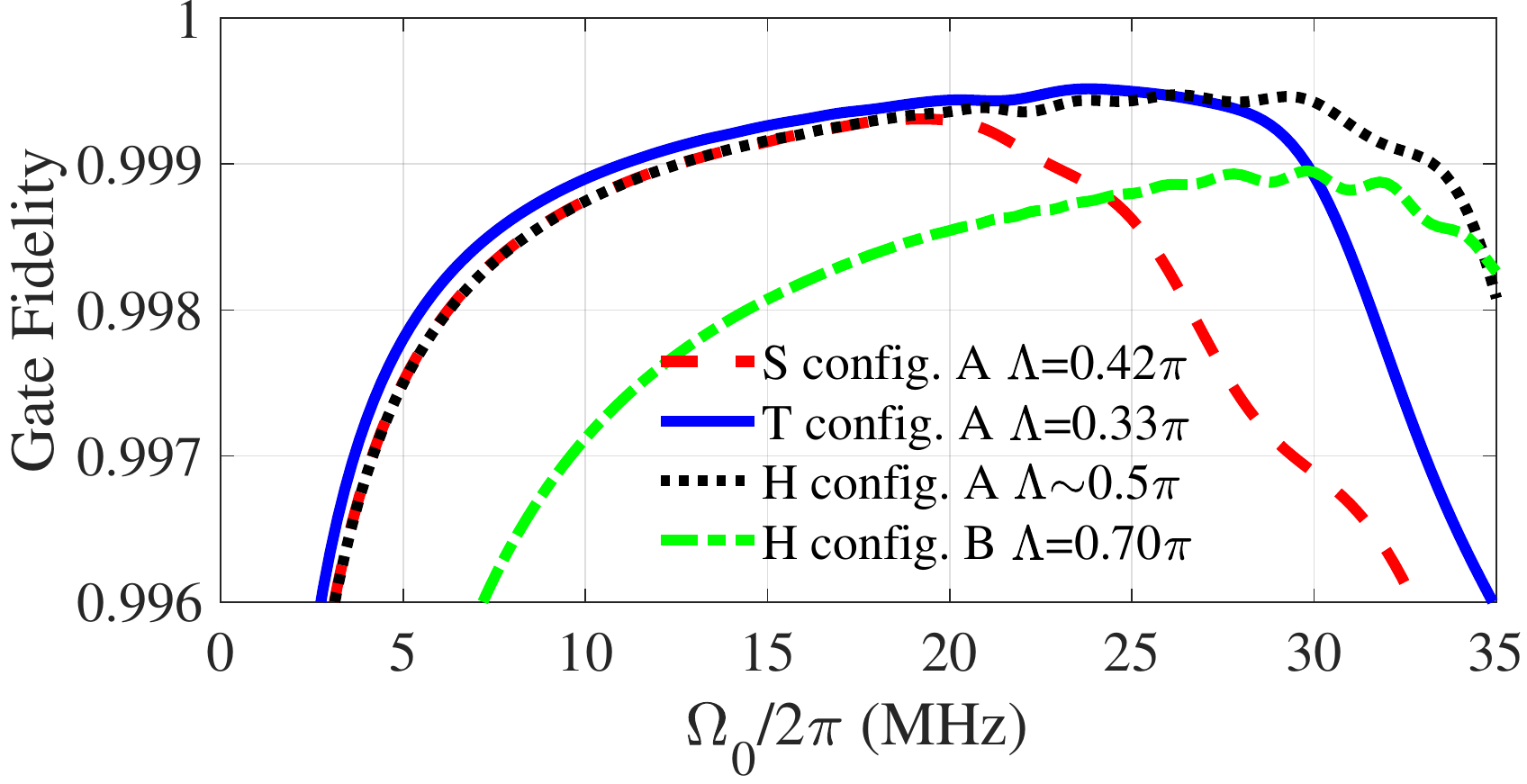}
\caption{The gate fidelities as a function of the pulse peak $\Omega_0$ for geometric S, T and H gates with their optimal path parameters (S gate: $\Lambda=0.42\pi$, T gate: $\Lambda=0.33\pi$, H gate: $\Lambda\simeq0.5\pi$), which has  the highest fidelities in the configuration A. For the configuration B, the optimal path parameter for the geometric H gate is $\Lambda=0.70\pi$. }\label{Figure5}
\end{figure}

\begin{figure}[tbp]
\flushright
\includegraphics[width=0.85\textwidth]{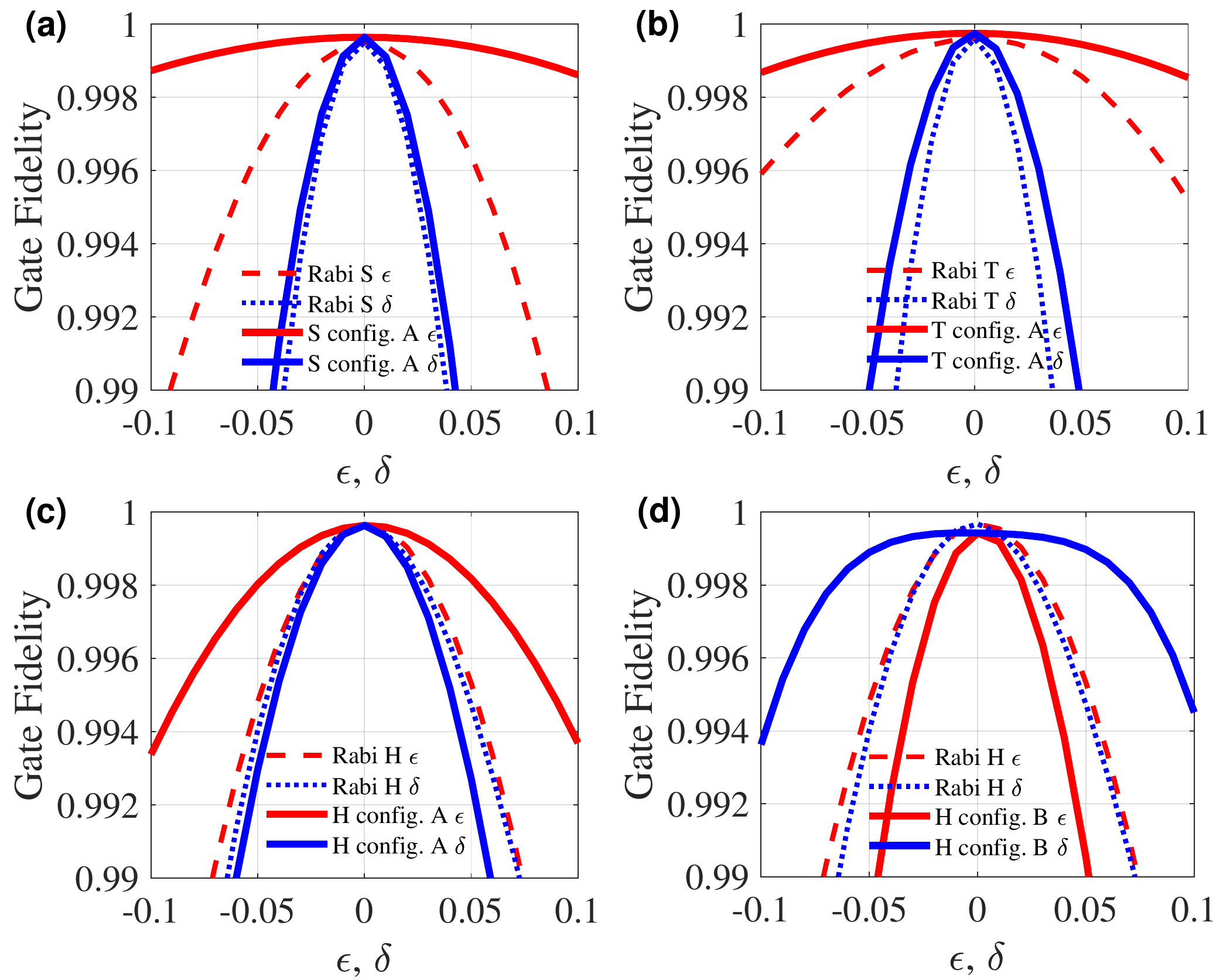}
\caption{Comparison of robustness between the geometric and corresponding dynamical gates against two local errors. (a-c) For the configuration A, gate-robustness for geometric S, T and H gates  with the optimal path parameters (S gate: $\Lambda=0.42\pi$, T gate: $\Lambda=0.33\pi$ and H gate: $\Lambda\simeq0.5\pi$) (d) The gate-robustness for geometric H gate with the optimal path parameter $\Lambda=0.70\pi$  in the configuration B.}\label{Figure6}
\end{figure}

The gate fidelity is defined as $F_1^{G}=\frac{1}{2\pi}\int^{2\pi}_0\langle\varphi_1^{f}|\rho_1|\varphi_1^{f}\rangle d\theta_1$ \cite{Poyatos} with the ideal final state  of $|\varphi_1^{f}\rangle=U_1|\varphi_1^{i}\rangle$, the integration is numerically done for $1001$ different initial states of $|\varphi_1^{i}\rangle=\cos \theta_1|0\rangle+\sin\theta_1|1\rangle$ with $\theta_1$ being uniformly distributed within the range of $[0, 2\pi]$. Besides, for stronger driving field, faster gate operation can be induced, and thus leads to less decoherence induced gate error. However, stronger driving field also causes more leakage to the non-computational subspace. Therefore, in figure \ref{Figure5}, we plot the gate fidelities as a function of the pulse peak $\Omega_0$ for geometric S, T and H gates with their optimal path parameters, i.e.,  $\Lambda=(0.42, 0.33,  0.5) \pi$,  where they have the highest fidelities in the configuration   A. Meanwhile, for geometric H gate from configuration  B,  the optimal path parameter is $\Lambda=0.70\pi$, which  has the strongest robustness against $\sigma_z$ error. Moreover, in the simulation, we have used a corrected shape $\Omega_D(t)=\Omega(t)-\{i\dot{\Omega}(t)+[\dot{\phi}(t)+\Delta]\Omega(t)\}/(2\alpha)$ for the driving field with the original simple pulse shape being $\Omega(t)=\Omega_0\sin^{2}(\pi t/T)$,  as it can be arbitrary. In addition, the decoherence rates and anharmonicity of the transmon \cite{Barends,Chen3} are $\kappa_{-}=\kappa_z=2\pi\times4$ kHz and $\alpha=2\pi\times220$ MHz, respectively.  By these setting, the maximal fidelities of S, T and H gates can be obtained as $F^{G}_{S}\simeq99.93\%$, $F^{G}_{T}\simeq99.95\%$ and $F^{G}_{H}\simeq99.95\%$ ($\simeq99.90\%$ for configuration B), and their corresponding uncorrected pulse peaks are $\Omega_0/(2\pi)=(19, 24, 30)$ MHz, respectively. Finally, based on the optimal pulse peaks and path parameters above-mentioned, as shown in figure \ref{Figure6}, we also plot the comparison of gate robustness between  geometric gates and their corresponding dynamical ones against two local errors,  in the form of $(1+\epsilon)\Omega(t)$ and $\Delta+\delta\Omega_0$.

\section{Nontrivial two-qubit geometric gates of path optimization}
The above path-optimized method can be readily generalized to the case of nontrivial two-qubit geometric gates. As shown in figure \ref{Figure1}(b), we here consider a two-dimensional lattice of  adjacently capacitive coupled transmon qubits, and the coupling strength there is usually fixed once the chip is fabricated. To realize tunable coupling \cite{Roth,Li,Cai,Chu} between two qubits, e.g., $T_1$ and $T_2$, we add an ac driving of $\dot{F}(t)$ with $F(t)=-\beta\cos{[\nu t+\varphi(t)]}$ on the qubit $T_2$ so that $\omega_2(t)=\omega_2+\dot{F}(t)$, and the interaction Hamiltonian can be calculated as
\begin{eqnarray}
\mathcal{H}_{12}(t)=g_{12}\Big\{\big[|10\rangle_{12}\langle01|e^{i\Delta_1t} +\sqrt{2}|11\rangle_{12}\langle02|e^{i(\Delta_1+\alpha_2)t}   \nonumber\\
\qquad\quad\quad+\sqrt{2}|20\rangle_{12}\langle11|e^{i(\Delta_1-\alpha_1)t}\big]e^{i\beta\cos{[\nu t+\varphi(t)]}}+\textup{H.c}\Big\},
\end{eqnarray}
where $\Delta_1=\omega_1-\omega_2$ with $\omega_1$ and $\omega_2$ being the frequencies of qubits $T_1$ and $T_2$ respectively, $g_{12}$ is coupling constant, and $\alpha_1$ and $\alpha_2$ are the intrinsic anharmonicities of the two qubits. Making a representation transformation $U=\exp\left\{-i\Delta'(|11\rangle_{12}\langle11|-|02\rangle_{12}\langle02|)t/2\right\}$ on $\mathcal{H}_{12}(t)$, and Hamiltonian $\mathcal{H}_{12}(t)$  reduces to
\begin{eqnarray}\label{eq_twoqubit}
\mathcal{H}'_{12}(t)=-\frac{1}{2}\left\{\Delta'\cdot\sigma'_z-g'_{12} \left[\cos{\varsigma(t)}\sigma'_x+\sin{\varsigma(t)}\sigma'_y\right]\right\},
\end{eqnarray}
where $g'_{12}\!=\!2\sqrt{2} g_{12}J_1(\beta)$, $\Delta'=\nu-\Delta_1-\alpha_2$, $\varsigma(t)\!=\!\varphi(t)-\pi/2$,   and $\{\sigma'_x,\sigma'_y,\sigma'_z\}$ are the Pauli operators represented by the subspace \{$|11\rangle_{12},|02\rangle_{12}$\}.  This represents a detuned interaction in the subspace of \{$|11\rangle_{12},|02\rangle_{12}$\}, and whose structure of energy levels is plotted in figure \ref{Figure1}(c). Note that we have utilized the Jacobi-Anger identity $\exp[i\beta\cos(\nu t+\varphi(t))] =\sum_{m=-\infty}^{\infty}i^mJ_m(\beta)\exp[im(\nu t+\varphi(t))]$ in obtaining  equation (\ref{eq_twoqubit}). Obviously,  Hamiltonian $\mathcal{H}'_{12}(t)$ has the same structure  as that of Hamiltonian $\mathcal{H}_1(t)$ in the single-qubit case. Therefore, we can construct arbitrary single-qubit-like geometric gates in the subspace \{$|11\rangle_{12},|02\rangle_{12}$\} through setting appropriate parameters, which corresponds to a two-qubit geometric control-phase gate in the two-qubit computational subspace $\{|00\rangle_{12},|01\rangle_{12},|10\rangle_{12},|11\rangle_{12}\}$, i.e.,
\begin{eqnarray}
U_2(\tau')=|00\rangle_{12}\langle00|+|01\rangle_{12}\langle01| +|10\rangle_{12}\langle10|+e^{i\gamma''_g}|11\rangle_{12}\langle11|.
\end{eqnarray}

 \begin{figure}[tbp]
\flushright
  \includegraphics[width=0.85\textwidth]{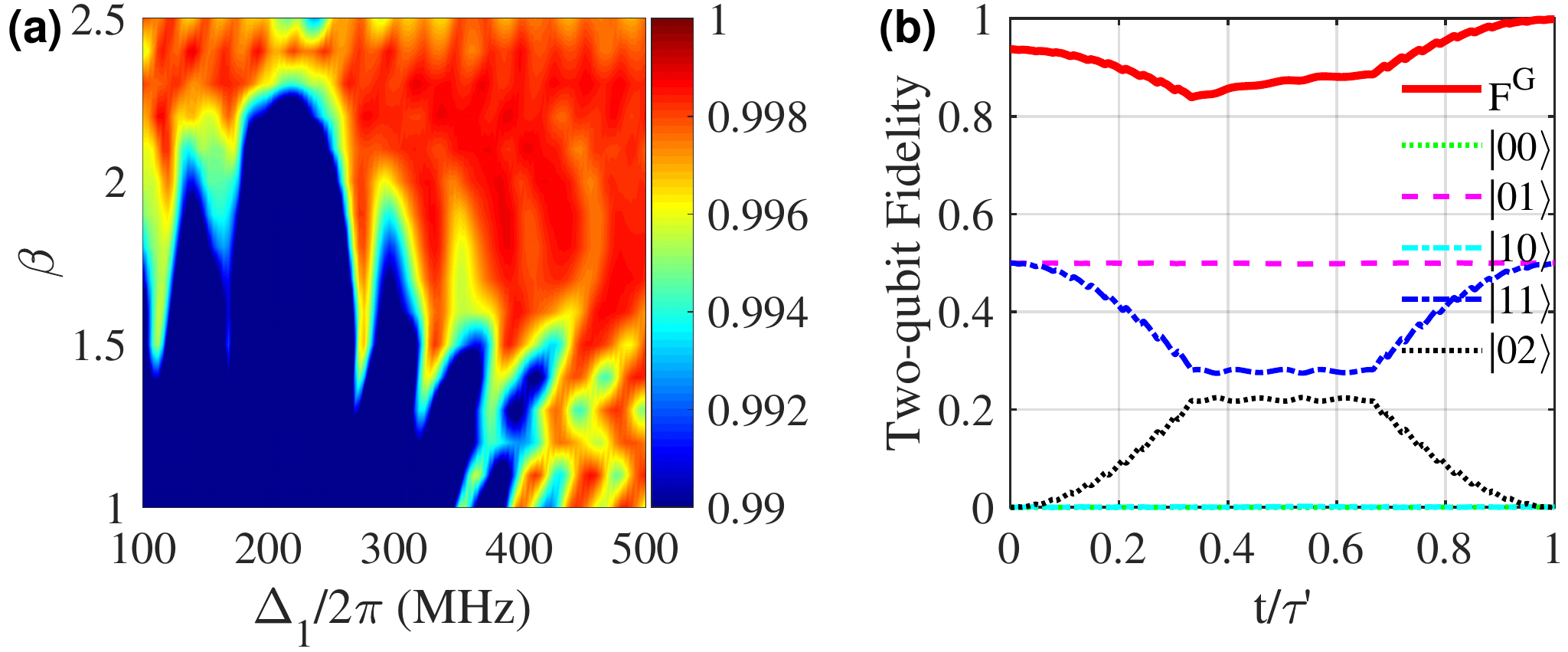}
\caption{(a) The gate fidelities of the implemented geometric control-phase gate ($\gamma''_g=-\pi/4$) as function of the parameters $\Delta_1$ and $\beta$ in the relative optimal path $\Lambda=0.45\pi$. (b) The graph for corresponding state populations with the inial state $(|01\rangle+|11\rangle)/\sqrt{2}$ and gate fidelity alter with the evolution time $t$ in the same path as the (a).}\label{Figure7}
\end{figure}

In the following, we set $\gamma''_g=-\pi/4$ as a typical example to test this gate performance, whose path can be optimized similar to the  single-qubit case. Then, the master equation can be rewritten as
\begin{eqnarray}\label{master2}
\dot{\rho}_2=-i[\mathcal{H}_{12}(t),\rho_2] +\frac{1}{2}\sum^2_{j=1}\left\{\kappa^j_z\mathcal{L}(\mathcal{A}^j_z) +\kappa^j_{-}\mathcal{L}(\mathcal{A}^j_{-})\right\},
\end{eqnarray}
where $\mathcal{A}^j_{-}=|0\rangle_{j}\langle1|+\sqrt{2}|1\rangle_{j}\langle2|$ and $\mathcal{A}^j_z=|1\rangle_{j}\langle1|+2|2\rangle_{j}\langle2|$. Besides, the two-qubit gate fidelity can be defined as $F_2^{G}=\frac{1}{4\pi^2}\int^{2\pi}_0 \int^{2\pi}_0 \langle\varphi_2^{f}|\rho_2|\varphi_2^{f}\rangle d\theta_1 d\theta_2$ for the ideal final state $|\varphi_2^{f}\rangle=U_2|\varphi_2^{i}\rangle$ with a general initial state $|\varphi_2^{i}\rangle=(\cos \theta_1|0\rangle+\sin\theta_1|1\rangle)\otimes(\cos \theta_2|0\rangle+\sin\theta_2|1\rangle)$, in which the integration is numerically performed for $10001$ input states with $\theta_1$ and $\theta_2$ uniformly  distributed in $[0, 2\pi]$. We set the relative optimal path $\Lambda=0.45\pi$ and plot the gate fidelities as a function of parameters $\Delta_1$ and $\beta$,  as shown in  figure \ref{Figure7}(a). We find that the two-qubit gate fidelities can exceed $99.85\%$ when the parameters $\beta$ and $\Delta_1$ are around $1.9\pm0.1$ and $2\pi\times(388\pm5)$ MHz, respectively. Furthermore, in the figure \ref{Figure7}(b), we plot the corresponding state populations and the gate fidelity dynamics with the inial state being $(|01\rangle+|11\rangle)/\sqrt{2}$, under the conditions of $\beta=1.9$ and $\Delta_1=2\pi\times388$ MHz, and the geometric control-phase gate fidelity can achieve as high as $99.87\%$. Moreover, in our numerical simulation,  all parameters are set to be readily accessible within current transmon experiment \cite{Kjaergaard}, such as the coupling strength $g_{12}=2\pi\times8$ MHz, transmon anharmonicities of $\alpha_1=\alpha_2=2\pi\times220$ MHz, $\kappa^1_{-}=\kappa^1_z=\kappa^2_{-}=\kappa^2_z=2\pi\times4$ kHz, and the detuning $\Delta' \simeq-2\pi\times83$ MHz. Note that, besides this high gate fidelity, the merit of gate robustness is similar as that of the case of single-qubit phase gate, as the structure of Hamiltonian $\mathcal{H}'_{12}(t)$ is the same as that of $\mathcal{H}_1(t)$.

\section{Conclusion}
In summary, we have proposed a path-optimized scheme for implementing nonadiabatic geometric gates on superconducting transmon qubits with simple and conventional experimental control, where the constructed single-qubit geometric S, T, and H gates are superior over the corresponding dynamical gates with regard to gate fidelity and robustness by selecting appropriate paths.  Through numerical simulation, we can get the fidelities for geometric S, T and H gates as $99.93\%$, $99.95\%$ and $99.95\%$, respectively.  Meanwhile, this method of path optimization can be readily generalized to the case of nontrivial two-qubit geometric gates, and the gate fidelity of geometric control-phase gate can reach $99.87\%$ within the available experimental parameters.  Remarkably, the gate robustness in our scheme is greatly enhanced besides the improvement of the gate-fidelity, which manifests the intrinsic robust nature  of the geometric phases. Finally, we do not set constrain on the pulse shapes in our construction of the geometric quantum gates, and thus  various pulse-shaping techniques can be potentially possible to be  incorporated into our scheme.  All of these merits  suggest that our scheme represents a new perspective for large-scale fault-tolerant solid-state quantum computation.

\section*{Acknowledgements}
This work was supported by the Key-Area Research and Development Program of GuangDong Province (Grant No.  2018B030326001), the National Natural Science Foundation of China (Grant No. 11874156), Guangdong Provincial Key Laboratory of Quantum Science and Engineering (Grant No. 2019B121203002), and Science and Technology Program of Guangzhou (Grant No. 2019050001).

\section*{Data availability statement}
The data that support the findings of this study are available upon reasonable request from the
authors.

\section*{Appendix A: The construction of dynamical gates}
From a two-level quantum system that is resonantly driven by a external field, the interaction Hamiltonian reads
\begin{eqnarray}
\mathcal{H}_d(t)=\frac{1}{2}\Omega(t)\left[\cos{\phi(t)}\sigma_x+\sin{\phi(t)}\sigma_y\right],
\end{eqnarray}
we set $\phi(t)=\phi_d$ as a constant, to ensure the geometric phase $\gamma_g\equiv0$. Thus, the dynamical evolution operator at the final time $\tau$ can be obtained as
\begin{eqnarray}\label{dynamical_unitary}
  U_d(\tau)=e^{-i\int_0^{\tau}\mathcal{H}_d(t)\,dt}=\sigma_z\cos\frac{\theta^d_{x,y}}{2} -i\sin\frac{\theta^d_{x,y}}{2}(\sigma_x\cos{\phi_d}+\sigma_y\sin{\phi_d}),
\end{eqnarray}
where $\theta^d_{x,y}\!\!=\!\!\int_0^{\tau}\Omega(t)dt$. When $\phi_d\!\!=\!\!0$ and $\pi/2$, the $X$- and $Y$-axis rotation operations $R_x^{d}(\theta^d_x)$ and $R_y^{d}(\theta^d_y)$ can be obtained, respectively, while the $Z$-axis rotation  can be obtianed by $R_z^{d}(\theta^d_z)\!\!=\!\!R_x^d(-\pi/2)R_y^d(-\theta^d_z)R_x^d(\pi/2)$. Specially, a set of universal single-qubit gates, i.e., Phase, $\pi/8$ and Hadamard gates can be realized as $R_x^d(-\pi/2)R_y^d(-\pi/2)R_x^d(\pi/2)$, $R_x^d(-\pi/2)R_y^d(-\pi/4)R_x^d(\pi/2)$ and $R_x^d(\pi)R_y^d(\pi/2)$, with total pulse areas $\mathcal{S}_d=\int_0^{\tau}\Omega(t)\mathrm{d}t=3\pi/2$, $5\pi/4$ and $3\pi/2$, respectively.

 \begin{figure}[tbp]
\flushright
\includegraphics[width=0.85\textwidth]{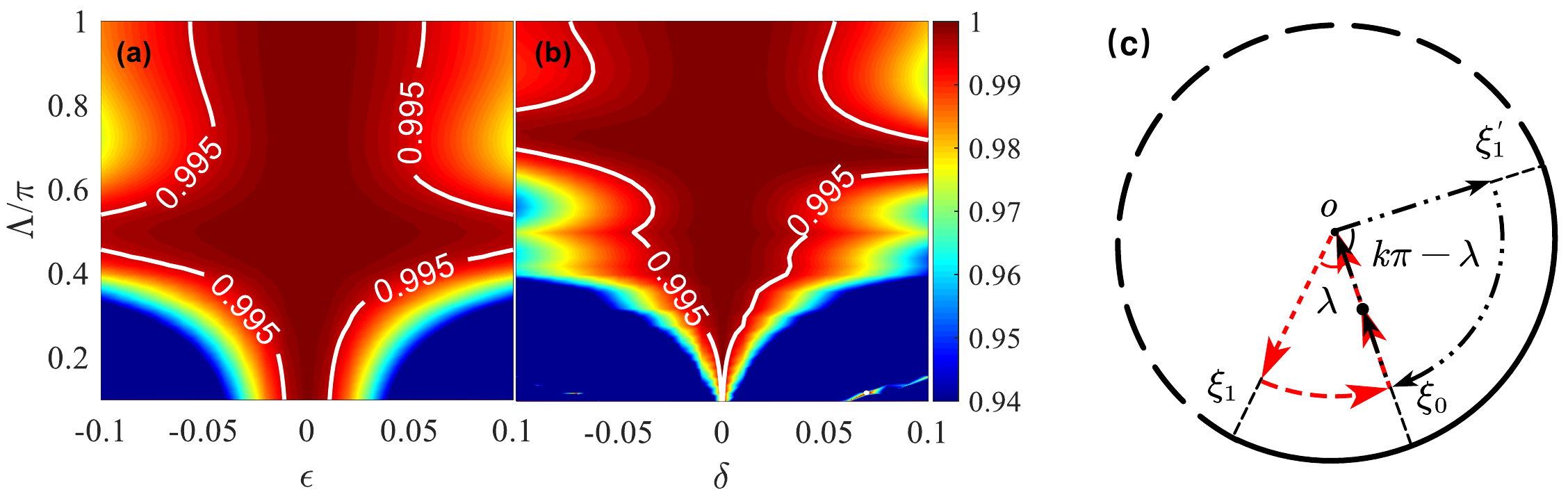}
\caption{Gate fidelties of the implemented geometric H gate for configuration B as a function of path parameters $\Lambda$ for (a) $\sigma_x$ and (b) $\sigma_z$ errors without decoherence, respectively, with $\epsilon,\delta\in[-0.1,0.1]$. (c) The vertical view of evolution trajectories in the Bloch sphere for configuration B (dash-dotted-black line) in contrast with the configuration A (dashed-red line).
  }\label{Figure8}
\end{figure}

\section*{Appendix B: Geometric quantum gates from configuration B}
Similar to those of the schemes in references \cite{yxu, chentao}, there is also a path configuration B versus configuration A for our scheme. But the difference is that, in our scheme, the path configurations can be further optimized by choosing different path parameters $\Lambda$, while the former one is only a special case of our scheme. Herein, in contrast to $\xi_1-\xi_0=\lambda$ in the configuration A, we can set $\xi'_1-\xi_0=\lambda-k\pi$ in the configuration B, which is shown in figure \ref{Figure8}(c). Then, the geometric phase $\gamma'_g$, in the configuration B, can be calculated as
\begin{eqnarray}
\gamma'_g=\frac{1}{2}(1-\cos\Lambda)(\xi'_1-\xi_0)=\gamma_g-\frac{k\pi(1-\cos\Lambda)}{2}.
\end{eqnarray}
In this case, we let $k\pi(1-\cos\Lambda)/2=\pi$ and have $k=2/(1-\cos\Lambda)$. Specially, when $\Lambda=\pi$, then $\xi'_1-\xi_0=\lambda-\pi$, which reduces to the configuration B for the conventional nonadiabatic GQC  \cite{yxu, chentao}. As a result, the new evolution operator is $U'_1(\tau)=\exp(i\gamma'_g\vec{n}\cdot\vec{\sigma})=-\exp(i\gamma_g\vec{n}\cdot\vec{\sigma})$ that is equivalent to $U_1(\tau)$ in the configuration A. Therefore, similar to the configuration A case, we can also optimize geometric quantum gates induced from configuration B, just changing $\xi_1$ to $\xi'_1$ in the path parameter.

\begin{figure}[tbp]
\flushright
\includegraphics[width=0.84\textwidth]{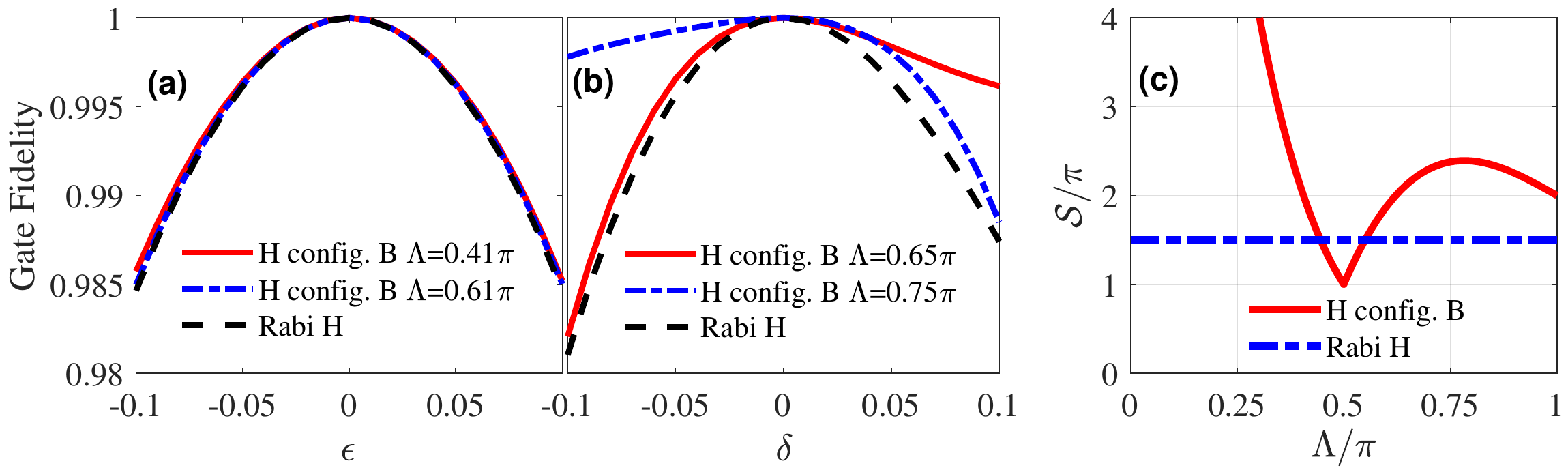}
\caption{Comparison of robustness between geometric H gate and corresponding dynamical gate (also denoted as ``Rabi gate'') against (a) $\sigma_x$ and (b) $\sigma_z$ errors for the boundary path parameters, respectively. (c) The comparison of the total pulse areas $\mathcal{S}$ for geometric H gate in the configuration B case with different path parameters and the corresponding dynamical gate.
  }\label{Figure9}
\end{figure}

For the $\sigma_z$ error, in the configuration A case, as shown in the maintext, there is no range of the path parameter that makes the robustness of geometric H gate is better than that of the corresponding dynamical gate. Here, we only consider the robustness of path-optimized geometric H gate in the configuration B. As shown in figures \ref{Figure8}(a) and \ref{Figure8}(b), we can clearly obtain the gate robustness for geometric H gate is distinct within the range of $\Lambda/\pi\in[0.1,0.5)\cup(0.5,1]$ under two different local errors $\epsilon,\delta\in[-0.1,0.1]$. Moreover, in figures \ref{Figure9}(a) and \ref{Figure9}(b), we compare the robustness of this gate with its corresponding dynamical gate under several boundary paths. We find that the geometric quantum gates possess better  robustness when the path parameter is within the ranges  of $\Lambda/\pi\in[0.41,0.5)\cup(0.5,0.61]$ and $\Lambda/\pi\in[0.65,0.75]$ for $\sigma_x$ and $\sigma_z$ errors, respectively. Obviously, it can be found that the stronger robustness against $\sigma_z$ error is, the weaker robustness against $\sigma_x$ error will be, which is just opposite to the result in the configuration A case. In addition, in the case of the configuration B, the pulse areas for geometric and dynamical H gate as a function of the path parameter  $\Lambda/\pi\in(0,0.5)\cup(0.5,1]$ are plotted  in figure \ref{Figure9}(c).

\end{document}